# A review of quantum collision dynamics in Debye plasmas

R. K. Janev,[1] Song Bin Zhang,[2,*] and Jian Guo Wang [3]

[1]Macedonian Academy of Sciences and Arts, P.O. Box 428, 1000 Skopje, Macedonia

[2]School of Physics and Information Technology, Shaanxi Normal University, Xi'an 710062, China

[3]Key Laboratory of Computational Physics, Institute of Applied Physics and Computational Mathematics, P.O. Box 8009, Beijing 100088, China

**Abstract**

Hot, dense plasmas exhibit screened Coulomb interactions, resulting from the collective effects of correlated many-particle interactions. In the lowest particle correlation order (pair-wise correlations), the interaction between charged plasma particles reduces to the Debye-Hückel (Yukawa-type) potential, characterized by the Debye screening length $D$. Due to the importance of Coulomb interaction screening in dense laboratory and astrophysical plasmas, hundreds of theoretical investigations have been carried out in the past few decades on the plasma screening effects on the electronic structure of atoms and their collision processes employing the Debye-Hückel screening model. The present article aims at providing a comprehensive review of the recent studies in atomic physics in Debye plasmas. Specifically, the work on atomic electronic structure, photon excitation and ionization, electron/positron impact excitation and ionization, and excitation, ionization and charge transfer of ion-atom/ion collisions will be reviewed.

## I. Introduction

The study of Coulomb interaction screening in plasma environments is one of the major subjects in plasma physics [1-6]. The Coulomb interaction screening in plasma environments is a collective effect of correlated many-particle interactions [7-9]. It strongly affects the electronic structure (spectral) properties of atoms and the properties of their collision processes with respect to those for isolated systems. Indeed, it has been observed experimentally in a number of laser-produced dense plasmas that the atomic spectral lines are significantly redshifted [10-14]. Note that the Debye-Hückel screening of Coulomb interaction between charged particles also appears in electrolytes, solid-state matter and many other physical systems (in nuclear physics it is known as Yukawa potential).

---

[*] song-bin.zhang@snnu.edu.cn

Extensive studies have been performed on the screening effects in classical hot, dense plasmas in the past decades (see, e.g., [7, 8] and references therein). These studies have been motivated mainly by the research in laser produced plasmas, EUV and X-ray laser developments, inertial confinement fusion and astrophysics (stellar atmospheres and interiors). The densities ($n$) and temperatures ($T$) in these plasmas span the ranges $n \sim 10^{15}$-$10^{18}$ cm$^{-3}$, $T \sim 0.5$-$5$ eV (stellar atmospheres), $n \sim 10^{19}$-$10^{21}$ cm$^{-3}$, $T \sim 50$-$300$ eV (laser produced plasmas) and $n \sim 10^{22}$-$10^{26}$ cm$^{-3}$, $T \sim 0.5$-$10$ keV (inertial confinement fusion plasmas). In classical hot, dense plasmas, both Coulomb and thermal effects play important roles. The relative importance of these two effects can be estimated by the so-called coupling parameter $\Gamma = \frac{\langle Z_i e \rangle^2}{R_i k_B T_e}$, where $\langle Z_i e \rangle$ is the average charge of ions in the plasma, $R_i = (\frac{3}{4\pi n_e})^{1/3}$ is the average inter-ionic distance, $k_B$ is the Boltzmann constant, $T_e$ and $n_e$ are the plasma electron temperature and density, respectively [4]. In the weakly coupled plasmas with relatively high temperatures and low densities, such as those created by laser irradiation of solids, met in the inertial confinement fusion research or in the stellar interiors, the potential energy is relatively small compared to the kinetic energy, long-range self-consistent interactions (described by the Poisson equation) dominate over short-range two-particle interactions (collisions) and $\Gamma \ll 1$. To the lowest particle correlation order (pair-wise correlations) the complete screened Coulomb potential in a more general way is given by [5, 6, 15, 16]

$$V(r) = \begin{cases} -Ze^2(\frac{1}{r} - \frac{1}{D+D_A}), & r \leq D_A \\ -Ze^2 \frac{D}{D+D_A} \frac{1}{r} e^{-\frac{r-D_A}{D}}, & r \geq D_A \end{cases}, \qquad (1)$$

where $Z$ is the nuclear charge, $D = \frac{k_B T_e}{(4\pi e^2 n_e)^{1/2}}$ and $D_A$ are the screening length and the mean minimum radius of the ion atmosphere, respectively. $D_A$ defines the ion sphere radius that the potential outside the ion sphere is screened by the plasmas, and $D_A < D$. In the limit when $D_A \to 0$,

Eq. (1) reduces to the most often used Debye-Hückel (Yukawa-type) potential [7, 8, 17]

$$V(r) = -\frac{Ze^2}{r} e^{-\frac{r}{D}} . \quad (2)$$

Alternatively, in strongly coupled plasmas with relatively low-temperature and high density ($\Gamma \gg 1$), the Coulomb effects are dominant (such as in the solid phase), the ions are packed tightly together; each ion occupies an equal volume and is surrounded by a sphere of radius $R_Z = [\frac{3(Z-1)}{4\pi n_e}]^{1/3}$ (the ion-sphere radius). Under these conditions the plasma screened Coulomb interaction is described by the ion sphere model potential, defined as [1, 2, 4, 17]

$$V(r) = \begin{cases} -\frac{Ze^2}{r}[1 - \frac{r}{2R_Z}(3 - \frac{r^2}{R_Z^2})], & r \leq R_Z \\ 0, & r > R_Z \end{cases} . \quad (3)$$

Note that while in the screened models Eq. (1) or Eq. (2) the thermal plasma effects dominate over the Coulomb effects, in the potential Eq. (3) the opposite is true; they obviously describe two different classical plasmas. More information about the models of these plasmas can be found in [1, 2, 4, 6, 7]. It should be noted that recently a modified Debye-Hückel potential [18-20] has been proposed to describe the interaction screening in dense quantum plasmas, where the de Broglie wavelength of the charge carriers is comparable to or larger than the inter-particle distance and plasma temperature is smaller than the Fermi temperature. Shukla and Eliasson [20] have shown that the effective potential of a test charge in a dense quantum plasma has the form of an exponential-cosine screened Coulomb potential

$$V(r) = -\frac{Ze^2}{r} e^{-k_q r/\sqrt{2}} \cos(k_q r/\sqrt{2}) , \quad (4)$$

where $k_e = [\frac{4m^2 \omega_p^2}{\hbar^2}]^{1/4}$ is the electron quantum wave number, $m$ is the electron mass, and $\omega_p = \sqrt{4\pi n e^2 m}$ is the electron plasma frequency. Usually quantum plasmas are characterized by a very low-temperature and a high number density. Such plasmas are met in metals, semiconductor devices, nanoscale structures (nanowires, quantum dots) and compact astrophysical objects (neutron stars, white dwarfs).

The above model potentials describe the interactions between the electron and the charged ion, while there are different arguments whether a similar Coulomb screening between two atomic electrons should be applied [1, 4, 15]. Generally, three type of models are employed in Debye plasmas in this respect: the first one does not consider any screening [21],

$$V_{ee}(r_1, r_2) = \frac{e^2}{|\mathbf{r}_1 - \mathbf{r}_2|}, \tag{5}$$

where $\mathbf{r}_1$ and $\mathbf{r}_2$ are the electron coordinates. The second one considers only the screening on one electron coordinate [4],

$$V_{ee}(r_1, r_2) = \frac{e^2}{|\mathbf{r}_1 - \mathbf{r}_2|} e^{-\frac{|\mathbf{r}_1|}{D}}. \tag{6}$$

The third one considers the screening on both coordinates [1],

$$V_{ee}(r_1, r_2) = \frac{e^2}{|\mathbf{r}_1 - \mathbf{r}_2|} e^{-\frac{|\mathbf{r}_1 - \mathbf{r}_2|}{D}}. \tag{7}$$

In most of the recent work the screening on both electron coordinates is taken.

In the present review, we shall provide a comprehensive overview of the fundamental theoretical studies of atomic physics in Debye plasmas modeled with screened interactions Eq. (2) and Eq. (7) in the past decade; earlier comprehensive reviews of hot-dense plasmas can be found in Refs. [1, 6-8]. In the sections below we summarize the work on atomic structure, photon collisions, electron collisions, positron collisions, and heavy particle collisions in a wide range of plasma screening conditions. Atomic units will be used in the remaining part of this article, unless explicitly indicated.

**II. Atomic structure**

In the nonrelativistic approximation, the radial Schrödinger equation for the hydrogenlike ion with nuclear charge $Z$ in screened potential Eq. (2), under the scaling transformations $\rho = Zr$, $\delta = ZD$, takes the form of that for the hydrogen atom

$$(-\frac{d^2}{2d\rho^2} + \frac{l(l+1)}{2\rho^2} - \frac{e^{-\rho/\delta}}{\rho}) P_{nl}(\rho, \delta) = \varepsilon_{nl}(\delta) P_{nl}(\rho, \delta), \tag{8}$$

where $\varepsilon_{nl}(\delta) = E_{nl}/Z^2$ and $P_{nl}(\rho, \delta)$ are the scaled energy and wave function, respectively. Thus, the results (energy levels, wave functions, photoionization cross sections) obtained for the hydrogen atom can be scaled to higher $Z$. The most prominent feature of the screened potential Eq. (2) is the lifting of the Coulomb $l$-degeneracy of the energy levels of hydrogenlike system (i.e., the energy of hydrogenic level $n$ splits into $n$ components). Another important feature of the potential Eq. (2) is that for any finite $\delta$, it supports only a finite number of bound $nl$ states. This implies that with decreasing $\delta$, the binding energies of $nl$ states decrease and the $nl$ energy levels successively enter in the continuum at certain critical screening lengths $\delta_{nl}$, obeying the relations $\delta_{n+1,l} > \delta_{nl}$ and $\delta_{n,l+1} > \delta_{nl}$. Table 1 shows the scaled critical screening lengths of hydrogenlike ions for the states with $n \leq 6$ [22]. Furthermore, with decreasing $\delta$, the excitation threshold energies also decreases. For a given $n$, the states with lower $l$ value have lower thresholds for any fixed value of $\delta$. As a consequence of the decrease of energies of bound states when $\delta$ decreases, the corresponding wave functions become increasingly more diffuse.

For the hydrogenlike system with large nuclear charge $Z$, the relativistic effect becomes important, fine structures of the energy levels and large and small components of the wave functions arise and should be studied with the Dirac equation [23]. By making the transformations $\rho = Zr$ and $\delta = ZD$, the radial wave functions of Dirac equation with the screened potential (2) is reduced to the scaled form

$$\begin{pmatrix} -\dfrac{e^{-\rho/\delta}}{\rho} - \varepsilon_{gk}(\xi,\delta) & \dfrac{1}{\xi}\left(\dfrac{k}{\rho} - \dfrac{d}{d\rho}\right) \\ \dfrac{1}{\xi}\left(\dfrac{k}{\rho} + \dfrac{d}{d\rho}\right) & -\dfrac{e^{-\rho/\delta}}{\rho} - \dfrac{2}{\xi^2} - \varepsilon_{gk}(\xi,\delta) \end{pmatrix} \begin{pmatrix} P_{gk}(\delta,\rho) \\ Q_{gk}(\delta,\rho) \end{pmatrix} = 0, \quad (9)$$

where $\xi = Z/c$, $k = \pm(j+1/2)$ for $l = j \pm 1/2$ is the relativistic momentum quantum number, $\varepsilon_{gk}(\xi,\delta)$ is the scaled energy. For bound states $g=n$, with $n$ being the principal quantum number, and for continuum states $g=\varepsilon$, with $\varepsilon$ being the kinetic energy of continuum electron. $P_{gk}(\delta,\rho)$ and $Q_{gk}(\delta,\rho)$ are the scaled large and small components of the electron wave function, whose scaled factors for the bound or continuum states are different. Eq.(9) also tells that, contrary to the nonrelativistic case, the results for $Z=1$ cannot be scaled to any higher $Z$ [24, 25]. Fig. 1 shows the scaled energies of $1s_{1/2}$, $2s_{1/2}$, $2p_{1/2;3/2}$, $3s_{1/2}$, $3p_{1/2;3/2}$ and $3d_{3/2;5/2}$ states of $Fe^{25+}$ as function of $\delta$, investigated by Xie et.al. [24]. As shown in the left panel, the energy splitting of $nl$ states increases with decreasing $\delta$, while the relativistic fine structure energy splitting of $nl$-states (noticeable in the figure for large values of $\delta$) decreases with decreasing $\delta$. The right panel illustrates that with increasing $\delta$, the scaled relativistic binding energies increase with increasing $Z$ and that the fine structure component of the $nl$ manifold with larger $j$-value has a smaller binding energy than the one with smaller $j$. The scaled critical screening lengths, $\delta_{nlj}$, at which the binding energies of $nlj$ state becomes zero, are 0.8343, 3.2048, 4.5047 and 4.5342 a.u. for $1s_{1/2}$, $2s_{1/2}$, $2p_{1/2}$ and $2p_{3/2}$ states, respectively. More information of the screening effects on the wave functions and phases of the continuum states can be found in the work of Xie et.al. [24].

For the three-body systems in Debye plasmas, significant attention has been paid to the screening effects on the resonant states (such as doubly excited states in two electron atomic systems or resonances in electron-atom scattering). These states play very important roles in the threshold electron dynamics, resulting in the dramatic changes of the threshold photoionization [26] and electron-impact excitation [27-29] and ionization [30, 31]. Ho and his associates have performed systematic and comprehensive investigations of doubly excited states or resonances of the typical three-body systems with the screening potentials, such as the hydrogen negative ion (H⁻) [32-36], the positronium negative ion (Ps⁻) [37-41], helium [42-44] and helium-like ions [45, 46]. We note that for many-electron atomic systems (with more than two electrons), most of the earlier works had incorporated the Debye screenings only in the electron-nucleus potential of Eq. (2) and Eq. (5), owing to the complicated derivation of the two-body screening potential and difficulties to perform the calculations [47]. But the screening effects on the valance electron dynamics of Li and Na are efficiently studied based on the model potential formalism [48-53].

The plasma screening effects on the polarizability is another active topic, since the polarizability is an important characteristic of an atomic or molecular system describing its response to an external electric field. Qi et. al. [54] systematically studied the static dipole polarizability of hydrogenlike ions in Debye plasmas. They found that with decreasing $\delta$, the contribution of the bound states to the polarizability decreases and that of continuum states increases. As a result, both the polarizabilities of 1s and 2s states gradually increase when $\delta$ decreases down to the critical screening length at which the 2p state merges into continuum, followed by a dramatic increase when 1s and 2s states become continuum states, respectively, after that only the continuum states contribute to the polarizability. Note that Ho and his associates also have studied the screening effects on the polarizabilities of hydrogen atom, H⁻, He and He-like ions [55-57]. Polarizabilities of Li and Na in Debye plasmas are also broadly investigated based on the model potential formalism [48, 51-53, 58, 59].

Spectroscopy is the most direct approach to study the screening effects on the atomic structure. Margenau et. al. [6] and Sil et. al. [60] have reviewed the spectroscopy in plasmas. Recently, the redshifts of atomic spectral lines have also been observed experimentally in a number of laser-produced dense plasmas [10-14]. Although there are many new relevant works [61-66], it is interesting to mention the work of Chang et. al. [66], when simulating the redshift of the Lyman-α emission line of H-like ions in plasmas, the calculations with a judicial choice of the radius of Debye sphere of the general Debye potential Eq. (1) generate the results in good agreement with the experimentally observed values, in addition to reproducing the simulated data consistent with more elaborate simulations based on quantum mechanical approaches.

**III. Photon collisions**

Studies of photo-excitation process in plasmas are mainly concentrated on the calculations of oscillator strengths [55, 65, 67-74], since the photo-excitation cross sections and radiative transition probabilities are directly related to the oscillator strengths [75]. Qi et. al. [65, 69] give a systematic presentation of the scaled spectral properties of hydrogenlike ions in Debye plasmas, including the transition frequencies, absorption oscillator strengths, radiative transition probabilities. The line intensities of the Lyman and Balmer series, are also presented in these references for a wide range of plasma screening parameters. It is shown that for the $\Delta n \neq 0$ transitions,the oscillator strengths and spectral line intensities decrease with increasing the plasma screening, while those for the $\Delta n=0$ transitions rapidly increase. The lines associated with the $\Delta n \neq 0$ transitions are redshifted, whereas those for $\Delta n=0$ transitions are blueshifted [65].

The plasma effects on the photoionization process have been studied in the past under various assumptions about the form of the screening defined by the plasma conditions [7, 8]. Studies of this process in a Debye plasma were reported in many papers [15, 22, 24, 26, 49, 50, 67, 68, 76-85]. The most prominent screening effects of the Debye plasmas on the energy behavior of photoionization cross sections of hydrogenlike ions are manifested in its low-energy region (Wigner threshold law, appearance of multiple shape and virtual-state resonances when the photoelectron energy is close to the bound or continuum energy of $nl$ states in the vicinity of their critical screening length, appearance of multiple Cooper minima associated with the $n,l+1$ shape resonances, (slight) reduction of the cross section at high photoelectron energies) [22]. As shown in Fig. 2 [24], when $\delta$ decreases to some critical values, the total scaled photoionization cross sections from the ground state of hydrogen atom and $Fe^{25+}$ ion in Debye plasmas are dominated by the contributions from shape resonances. Since relativistic effect is very important for $Fe^{25+}$ ion, the energy behavior and the magnitude of the scaled cross sections with the same $\delta$ for H and $Fe^{25+}$ behave differently. They are identical for the unscreened case, very close for $\delta$=20, 9, and 5 a.u., but quite different for other selected $\delta$. All observed differences and similarities between the cross sections in the figure for the same $\delta$ can be easily understood by taking into account the difference in the fine-structure energy splitting of bound states between the H and $Fe^{25+}$ ion and that all other considered values of $\delta$ lie in the vicinity of critical screening lengths at which $2p_{1/2;3/2}$ and $3p_{1/2;3/2}$ states merge into the continuums. Two or one resonance peaks appear in the photoionization of $Fe^{25+}$ for a given screening length depending on whether a shape resonance is formed in both $1s_{1/2} \rightarrow \varepsilon p_{1/2}$ and $1s_{1/2} \rightarrow \varepsilon p_{3/2}$ transitions or only in one of them. Note that in the case of H atom the fine-structure splitting is negligible, the critical screening lengths of $p_{1/2}$ and $p_{3/2}$ states coincide and so do the shape resonances ($p_{1/2;3/2}$) producing only one resonance peak in photoionization cross sections.

In many electron atomic systems, Feshbach resonances [86] dominate the photoionization cross sections in the low energy region. In such cases, the screening effects alter the properties of the resonances, resulting in the significant changes in the cross sections [15, 26, 84]. A typical example is the photodetachment of hydrogen negative ions in Debye plasmas [26], where the transformation of a Feshbach resonance into a shape resonance happens with the decrease of screening length, as shown in Fig. 3. Such transformation is manifested in the photoionization cross sections as change of the shape of the contributed peak from an "asymmetric" to a "symmetric". (A more detailed description of the crossover of Feshbach resonances to shape resonances is given the next section and in Refs. [28, 29]). Due to the softening of the screening potentials, the positions of the peaks or the resonances shift to the lower energies.

Another remarkable feature of the screening effects on the photoionization cross sections is the appearance of Cooper minima [87, 88]. No Cooper minima exist in the photoionization cross sections from 2s or 3s states of hydrogenlike ion and ground state of Li atom in the unscreened case. However, Cooper minima can appear in both of these two cases when the screening interactions increases to some extent [22, 49, 50, 79, 81, 82]. In the hydrogenlike ion in Debye plasmas, Cooper minima do not appear from the states whose radial wave functions do not have nodes, but Combet-Farnoux minima [89] are observed [22].

**IV. Electron collisions**

About 30 years ago (1980s), Weisheit et.al. [1, 4, 90] have studied the plasma screening effects on electron-impact excitation and ionization of hydrogenlike ions by the first Born approximation and close-coupling methods. In the studies, the screening interaction between the projectile electron and target electron was considered, but the changes of target wave functions and bound state energies were not taken into account. Later, Jung et. al. [91-95] have also investigated the plasma screening effects on electron-impact excitation and ionization processes in the Born approximation and the semiclassical impact parameter approximation, in which the plasma screening effects on both the bound and scattering electron and were considered. The variational method combined with the perturbation theory was applied to calculate the target bound states in the screened potential. These studies found that the plasma screened interaction significantly alters the electron-impact excitation/ionization cross sections.

It is well known that resonances play very important roles in electron-atom scattering and dominate the excitation cross sections in low energy (especially the near-threshold) region. Zhang et. al. [27-29] were the first to address the effects of screened Debye-Hückel interaction on the electron-atom scattering and excitation in energy region near the excitation threshold. The phenomenon of crossover of Feshbach resonances into shape-type resonances when the strength of the interaction screening varies was discovered. The specific studies were made for the electron-impact excitation of hydrogen atom in the energy region near the $n=2$ and $n=3$ excitation thresholds. The electron-proton and electron-electron screened Coulomb interactions were taken in the Debye-Hückel form (Eqs.(2) and (7), respectively) and the R-matrix method with pseudo states [96, 97] was used in scattering calculations. It was found that as the interaction screening increases, the $^{1,3}P$ and $^1D$ Feshbach resonances transform into shape-type resonances when they pass across the 2s and 2p threshold, respectively. As shown in Fig. 4, the widths of Feshbach resonances $^{1,3}S$, converging to the 2s threshold, rapidly decrease when the resonance approaches the threshold before it merges with the parent 2s state; while the widths of $^{1,3}P$ Feshbach resonances also considerably decrease when they approach the 2s threshold, but after passing it, their widths start to increase rapidly, a signature of the shape resonance [see the $D$ dependence of the $^1P^o(2)$ shape resonance in Fig. 4]. It is argued that this phenomenon results from the lifts of the $l$ degeneracy of $n=2$ Coulomb energy level by the screening interaction, and the changes of the main configurations of Feshbach resonances by the mixing of 2p state with higher $l$ states. The resonance transformation leads to dramatic effects in the 1s→1s, 1s→2s and 1s→2p excitation collision strengths in the n=2 threshold collision energy region, as shown in Fig. 5 where the dynamic evolution of 1s→2s collision strengths when the screening length varies is displayed. When the $^3P^o(2)$ and $^1P^o(1)$ resonances have already acquired a shape-type character, peaks are clearly observed in the 1s→2s collision strength for D=45 a.u. (at E=0.74794 Ry) and for D=29 a.u. (at E=0.745118 Ry), respectively. The effect of $^1D^e$ resonance on the 1s→2s collision strength is also observed after passing the 2s threshold at D=19 a.u. [28, 29]. Similar phenomena are also observed near the *n*=3 threshold, but the situation is more complex, since the threshold energy in the screened case is split into three components, with 3s, 3p, 3d energy levels having their own critical screening lengths [29]. Note that Kar and Ho [32-34, 41] have systematically studied the resonances in hydrogen negative ion with screened Coulomb interaction employing the highly accurate complex-coordinate rotation and the stabilization methods.

For high energy electron scatterings, the fast projectile electron is hardly affected by the (screened) interaction potentials, and can be well described by a plane wave; the excitation cross sections are directly related to the generalized oscillator strengths (GOS). However, the screened Coulomb interaction alters the bound state wavefunctions, resulting in changes of GOS and excitation cross sections. Qi et. al. [69] found that the plasma screening of the interaction reduces the GOS for transitions between the states with different *n* and increases the GOS between the states with the same *n*. The differential and total excitation cross sections are affected in a similar way when the strength of interaction screening varies..

Zammit et. al. [98-100] have performed comprehensive studies on the excitation and ionization processes in electron-hydrogen and electron-helium collisions in Debye plasmas employing the convergent close-coupling method [101] in the energy region from threshold to several hundreds of eV. They found that as the strength of the screening increases, the excitation cross sections decrease, while the total ionization cross section increases.

Qi et. al. [30, 31] also studied the fast-electron-impact ionization process of hydrogen-like ions in Debye plasmas. They considered the single differential ionization cross sections (SDCS) of hydrogen-like ions in their 2s and 2p initial states and focused on the low energy spectrum of ejected electrons. The SDCS of 2p state is at an impact electron energy of 1 keV/$Z^2$ shown in Fig. 6 for a number of scaled screening length δ=ZD as function of the scaled energy of ejected electron. The appearance of the sharp peaks in the SDCS for δ=10.88. 10.90, 10.22 a.u. is related to the fact that for these values of δ the 3d electron bound state is already in the continuum ($\delta_{3d}$=10.947 a.u., see Table 1) and the continuum εd electron is temporarily trapped by the centrifugal barrier of the effective potential and, thus, the ionization proceeds *via* the shape resonances in the effective potential, note that for δ=11.0 > $\delta_{3d}$ such peak is absent in the 2p SDCS. The SDCSs for δ=7.21 a.u. and δ=7.22 a.u. show an enhancement over a broader energy range of ejected electron. These two values of δ are in the immediate vicinity of the critical screening length of 3s bound state ($\delta_{3s}$=7.172 a.u.) indicating that the ionization process involves virtual intermediate states since for the s-continuum states there is no centrifugal barrier in the effective potential. The SDCSs for δ=8.85 a.u. and δ=8.89 a.u. show respectively a sharp peak and a broad enhancement in the low-energy region they are on the left and right side of the critical screening length of 3p state bound ($\delta_{3s}$=8.872a.u., cf. Table 1). The profiles for SDCS of fast-electron-impact ionization are similar to those of the photoionization cross sections of hydrogen-like ions in Debye plasmas [22, 85], except that the photoionization process involves only dipole transitions while the electron-impact ionization includes summation over all multi-pole transitions.

**V. Positron collisions**

Zhang et.al. [102] have studied positron-impact excitation of hydrogen atom in Debye plasmas by using the close-coupling method but without inclusion of the positronium formation channels. They found that the interaction screening decreases the coupling matrix elements, resulting in the significant reduction of excitation cross sections for 1s→2s, 1s→2p and 2s→2p transitions. This finding was supported by the more sophisticated treatment of Ghoshal et. al. [103, 104] employing the distorted-wave theory in the momentum space with inclusion of the positronium formation channels. Furthermore, the differential cross sections for the H(*ns*) →H(*nl*) elastic and inelastic transitions in both Debye and quantum plasmas have been also investigated by Ghoshal et. al. [103-107].

Positronium (Ps) formation in positron-hydrogen atom collisions in Debye plasmas is another active topic [107-110]. Sen et. al. [110] were the first to report positronium formation cross sections for positron-hydrogen atom collisions in Debye plasmas by using the second-order distorted-wave approximation. Later, Ma et. al. [109] have published Ps (n=1, 2) formation cross sections obtained by employing the momentum-space coupled-channel optical method [111]. As shown in Fig. 7 the Ps formation threshold energy decreases as the values of *D* decreases, since the binding energy of the atomic electron decreases as the Debye length decreases. The Ps formation cross sections are significantly larger (particularly in the threshold region) than that in the plasma free case. It can be observed from the figure that when the screening length decreases, the position of the maximum of Ps formation cross section shifts towards lower energies while the magnitude of the cross section maximum increases. It can also be observed in this figure that the plasma screening effect on the Ps formation cross section decreases as the projectile energy increases [109, 110]. We mention that Ghoshal et. al. [107, 108] have also studied the plasma screening effects on the differential cross sections of Ps formation in positron-hydrogen atom collisions. while Pamdey et. al. [112] have studied the Ps formation in positron-alkali-metal collisions in Debye plasmas based on the Debye screening of an electron-ion core model potential.

**VI. Heavy particle collisions**

The early studies involving heavy-particle collisions in hot, dense plasmas are those for proton-impact excitation of n=2 fine structure levels of hydrogen-like ions within a close-coupling scheme employing both the static Debye- Hückel and the ion-sphere model potentials [2], the electron capture in proton-hydrogenic ion collisions [113] and the symmetric the resonant charge exchange in hydrogen-like ion-parent nucleus collisions [114] by the classical Bohr-Lindhard model, and the classical trajectory Monte Carlo study of electron capture and ionization in hydrogen atom-fully stripped ion collisions [115]. However in those studies, the changes of the electronic structures (wave functions and energy levels) in the screened potential were taken into account at most within the first-order perturbation theory. Until recently, Wang and his associates performed nonperturbative comprehensive studies of the excitation, electron capture and ionization processes in Debye plasmas for $H^+$-H [116, 117], $He^{2+}$-H [118, 119], $He^{2+}$-$He^+$ [120], $C^{6+}$-H [121], $O^{6+}$+H [122], $N^{5+}$-H [123] and $O^{8+}$-H [124] collision systems by using the two-center atomic orbital close-coupling (TC-AOCC) method [125] in the intermediate energy region (1- 300 keV/u), and in $H^+$-H [126] and $He^+$-H [127] collisions by using the quantum-mechanical molecular orbital close-coupling (QMOCC) method [128] in the low energy region (below 1 keV/u) .

In the intermediate energy region, one typical work is the study of ionization in $He^{2+}$-H

collisions [118] by the TC-AOCC method. Fig. 8 shows the ionization cross sections to the target continuum (ITC) and to the projectile continuum (IPC) for different screening lengths in the energy range 5–300 keV/u [118]. With decreasing D, the ITC cross section first increases in the entire energy range considered down to D≤4 a.u., but then starts to decrease in the energy region above ~40 keV/u. This behavior can be attributed to the similar behavior of the direct coupling matrix elements. The IPC cross sections for the selected screening lengths have significant values only for energies below ~60 keV/u, and increase sharply with decreasing D. It can be understood from the fact that more and more bound states of $He^+$ become continuum states with decreasing D, and when D=2 a.u., the $2p(He^+)$ state, quasiresonantly coupled with the initial 1s(H) state, also becomes a quasicontinuum state, leading to a drastic increase of the IPC cross section with respect to the case of D =2.5 a.u..

It has been demonstrated in [129] that in $H^+$+H collisions the Regge poles of the scattering matrix are the physical origin of the oscillation structures in the elastic and electron capture cross sections in this collision system in the energy range 0.01–1.0 eV. Wu et. al. [126] have recently studied the in $H^+$+H collision in a Debye plasma and calculated the scattering matrix by using the QMOCC method. As shown in Fig. 9, they found that the number of Regge oscillations in the elastic and resonant charge transfer cross sections is quasi-conserved when the plasma Debye length $D$ is larger than 1.4 a.u., reflecting the invariance of the number of vibrational states of $H_2^+$ with changing $D$ in that region. Similarly, the frequency and amplitudes of glory oscillations in the elastic cross sections are quasi-invariant with the variation of $D$.

Note that in the high energy region, Pandey et. al. [129, 130] have studied the charge exchange and ionizaiton in $O^{8+}$-H and He-like system-H collisions in Debye plasmas by classical trajectory Monte Carlo method, and Bhattacharya et al. [131] have investigated the proton-hydrogen collisions in Debye plasmas by distorted wave formalism.

## VII. Summary

In conclusion, we have reviewed the recent studies of the screening effects of Debye plasmas on the atomic structure and collision processes. The plasma screening effects affect the atomic structure in several fundamental ways: reduction of the number of bound states, decreasing of the energy of bound states, broadening of the radial distributions of the bound states and changing the phase and amplitude of the continuum waves. All these changes drastically affect the dynamics of collision processes taking place in Debye plasmas, as discussed in the present article. The studies of the electronic structure of atoms and their collision processes in Debye plasmas in past few decades have revealed many new features of the screening effects on atomic physics and have contributed to a better understanding of the properties of these plasmas. The newly acquired knowledge should be useful in the simulation and diagnostics of hot, dense plasmas.

**Acknowledgement**

Thanks to L. Y. Xie (Fig. 1 and 2), Y. Y. Qi (Fig. 6), J. Ma (Fig. 7), L. Liu (Fig. 8) and Y. Wu (Fig. 9) for sending us the ASCII data of their published papers. S. B. Zhang was supported by the starting grants of Shaanxi Normal University (Grant Nos. 1112010214 and 1110010733). J. G. Wang was supported by the National Basic Research Program of China (Grant No. 2013CB922200).

**Table captions:**

Table 1. Values of the critical scaled screening length of hydrogenlike ions for the states with n≤6 [22].

**Figure captions:**

Fig. 1. (Color online) Scaled energies of 1*s*, 2*l*$_j$ and 3*l*$_j$ states of $Fe^{25+}$ (Z=26) ion as function of scaled screening length (left panel), and behavior of 2p$_{1/2}$, 2p$_{3/2}$, 3p$_{1/2}$ and 3p$_{3/2}$ energies near the critical screening lengths $\delta^c_{nlj}$ for the $Fe^{25+}$ (solid lines) and for hydrogen (Z=1) atom (dashed lines) (right panel) [24].

Fig. 2 (Color online) Scaled total photoionization cross sections for the ground $1s_{1/2}$ state of hydrogen (Z=1) atom (left panel) and $Fe^{25+}$ (Z=26) ion (right panel) as function of scaled

photoelectron energy for different scaled screening lengths [24].

Fig. 3 (Color online) Dynamic evolution of photodetachment cross sections around the n = 2 excitation threshold for different screening length [26]. $^1P^\circ(T)$ denotes the dominant resonance, where T = F (Feshbach) or S (Shape) resonances.

Fig. 4 (Color online) Variation of the widths of Feshbach and shape resonances when the screening length decreases [28, 29]. Short dashed lines represent the critical values of D where Feshbach resonances pass across the 2s or 2p threshold.

Fig. 5 (Color online) Dynamic evolution of 1s→2s collision strength with decreasing the Debye length [28, 29]. $^{2s+1}L^\pi(n)$ denotes the dominant resonance.

Fig. 6 (Color online) Electron-impact single differential cross sections from 2p state of hydrogenlike ion with incident scaled energy $\varepsilon_a$=1 keV [31].

Fig. 7 (Color online) Positronium (n=1 and n=2) formation cross sections in positron-hydrogen collisions for various Debye lengths [109].

Fig. 8 (Color online) Ionization cross sections to target continuum (ITC) states and to projectile continuum (IPC) states for different Debye lengths [118].

Fig. 9 (Color online) Regge cross section calculation for the $H^+$+H collision in the Debye plasma with D = 3.0 and 1.4 au [126]. (a) Regge trajectories in the energy range $0.00006 \leq E < 1$ eV for D=∞ (black solid line), D = 3.0 au (filled symbols) and D = 1.4 au, (hollow symbols). ((b) and (c)) Extracted Regge contribution and exact quantal charge transfer cross sections for D = 3.0 au and D = 1.4 au, respectively.

|   | *l* | | | | | |
|---|---|---|---|---|---|---|
| *n* | 0 | 1 | 2 | 3 | 4 | 5 |
| 1 | 0.839907 | | | | | |
| 2 | 3.222559 | 4.540956 | | | | |
| 3 | 7.171737 | 8.872221 | 10.947492 | | | |
| 4 | 12.686441 | 14.730720 | 17.210209 | 20.067784 | | |
| 5 | 19.770154 | 22.130652 | 24.984803 | 28.257063 | 31.904492 | |
| 6 | 28.427266 | 31.080167 | 34.285790 | 37.949735 | 42.018401 | 46.458584 |

Table 1

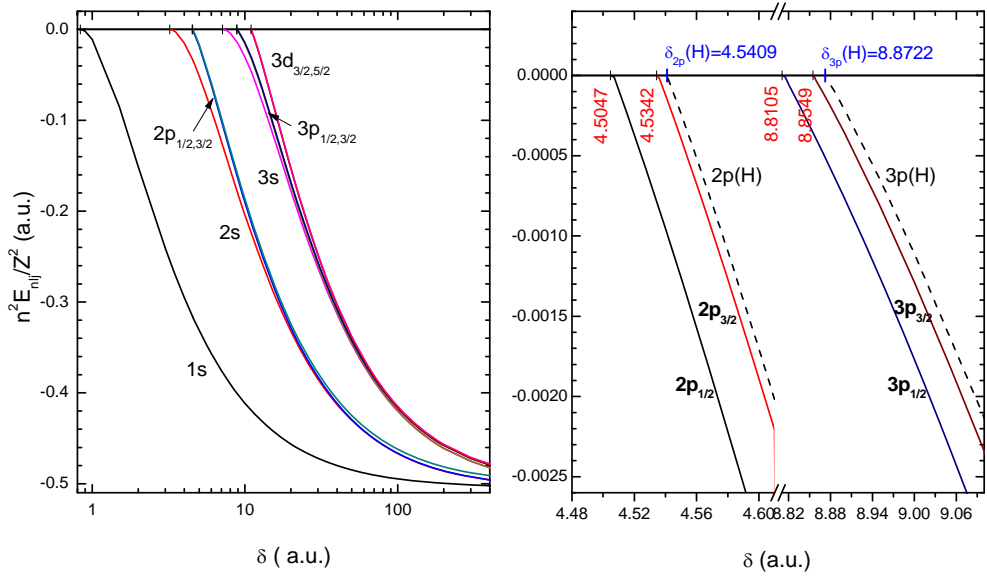

Fig. 1.

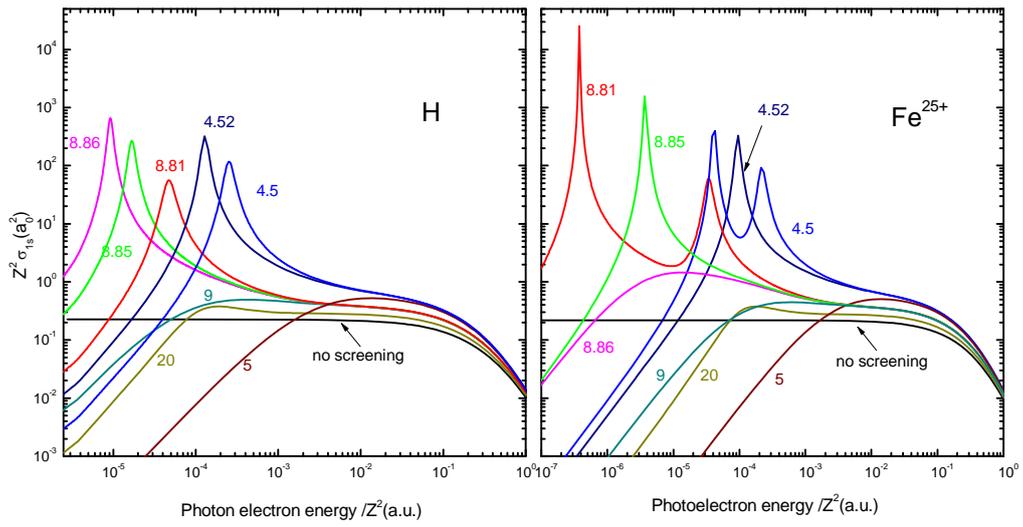

Fig. 2

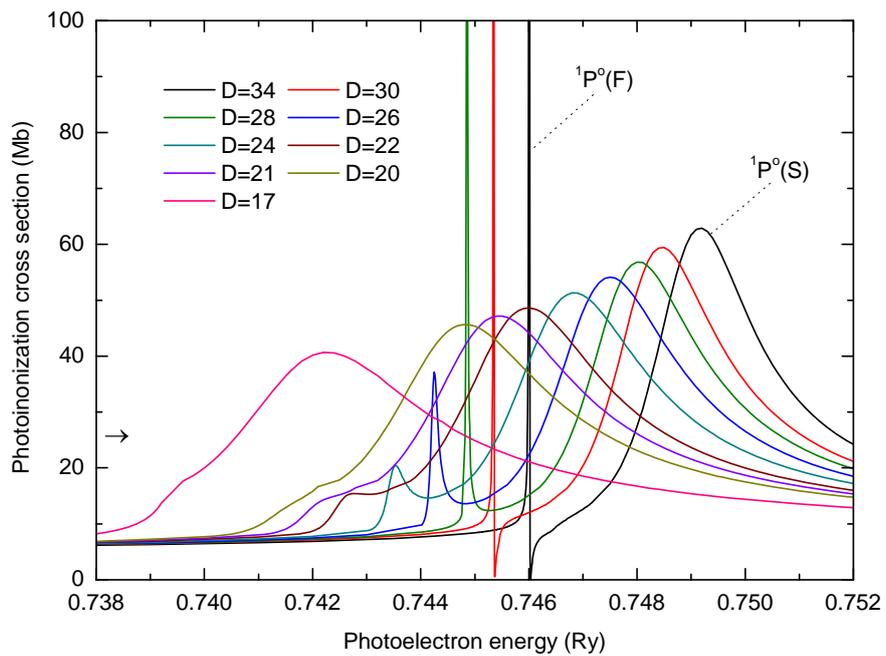

Fig. 3

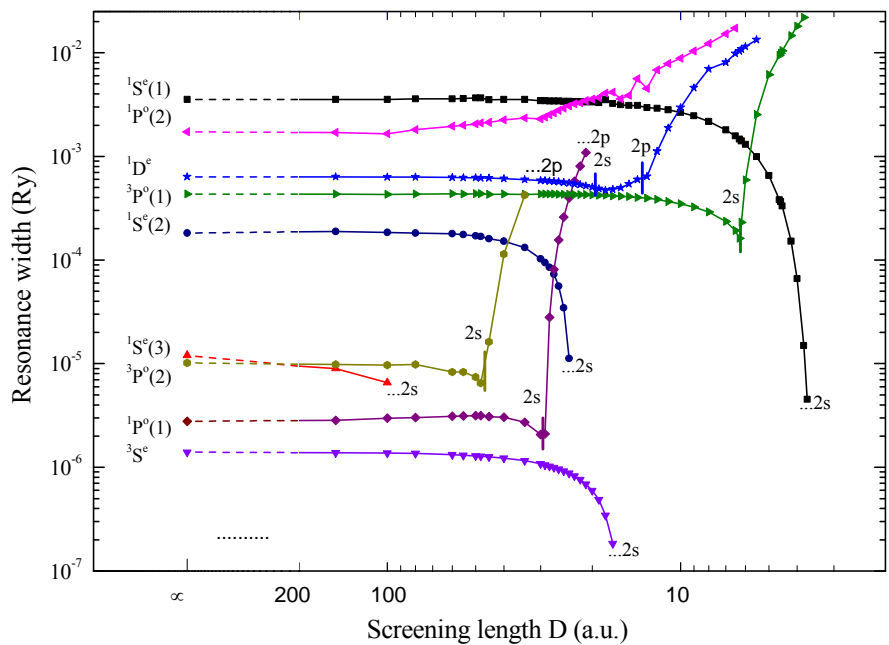

Fig. 4

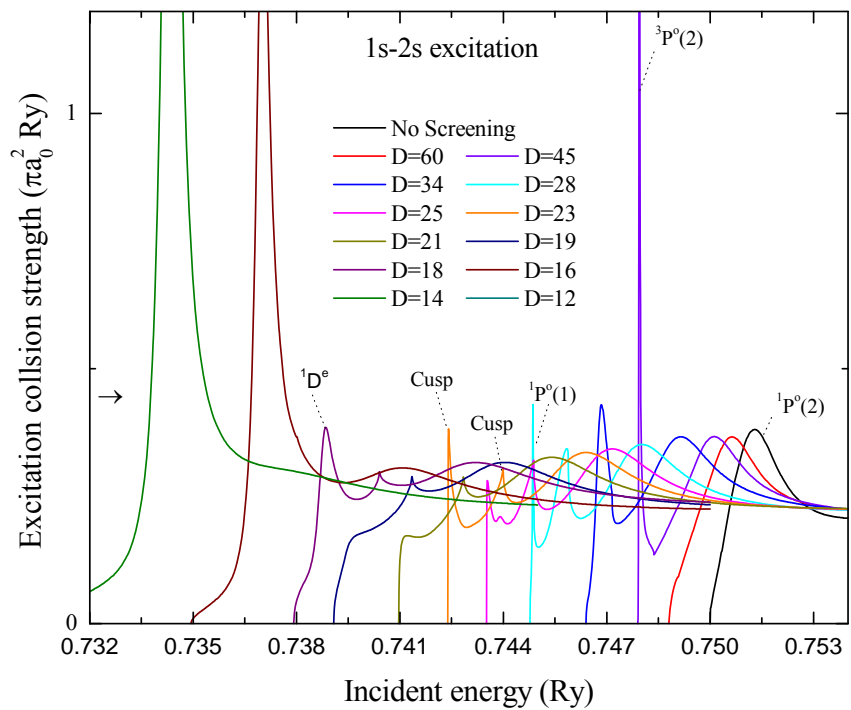

Fig. 5

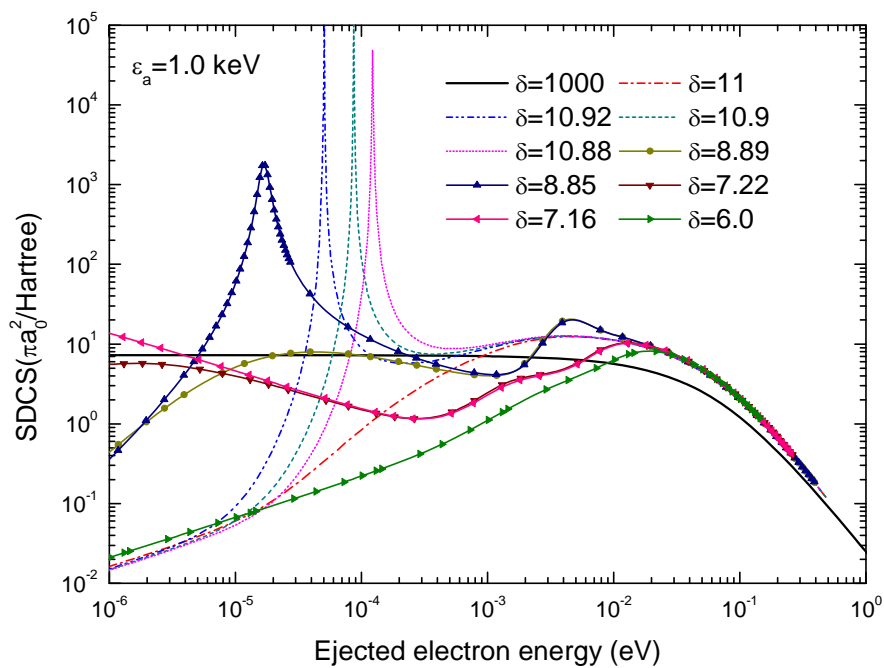

Fig. 6

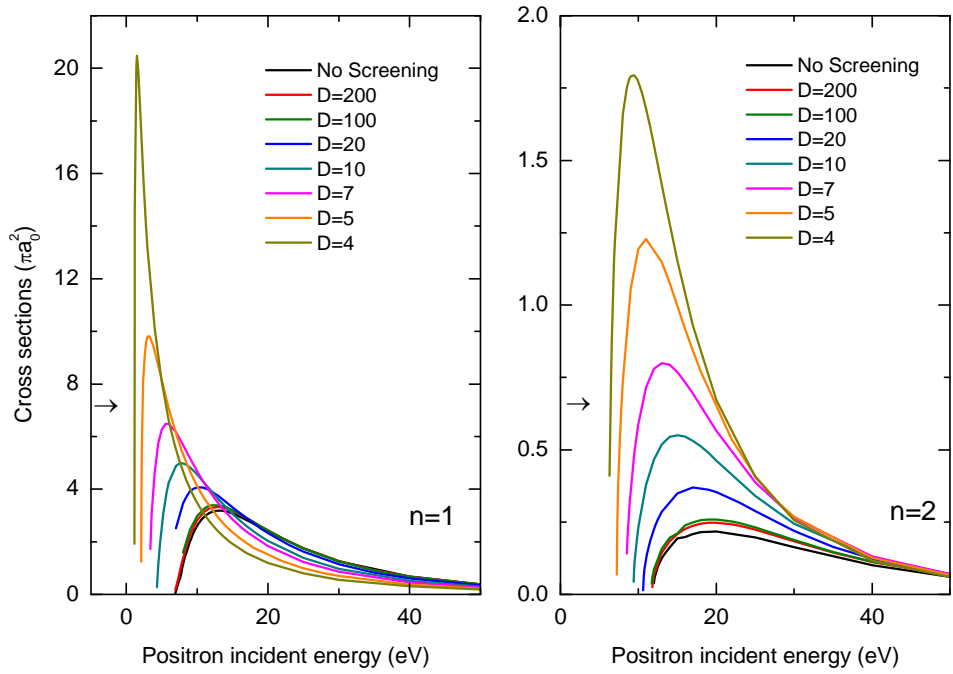

Fig. 7

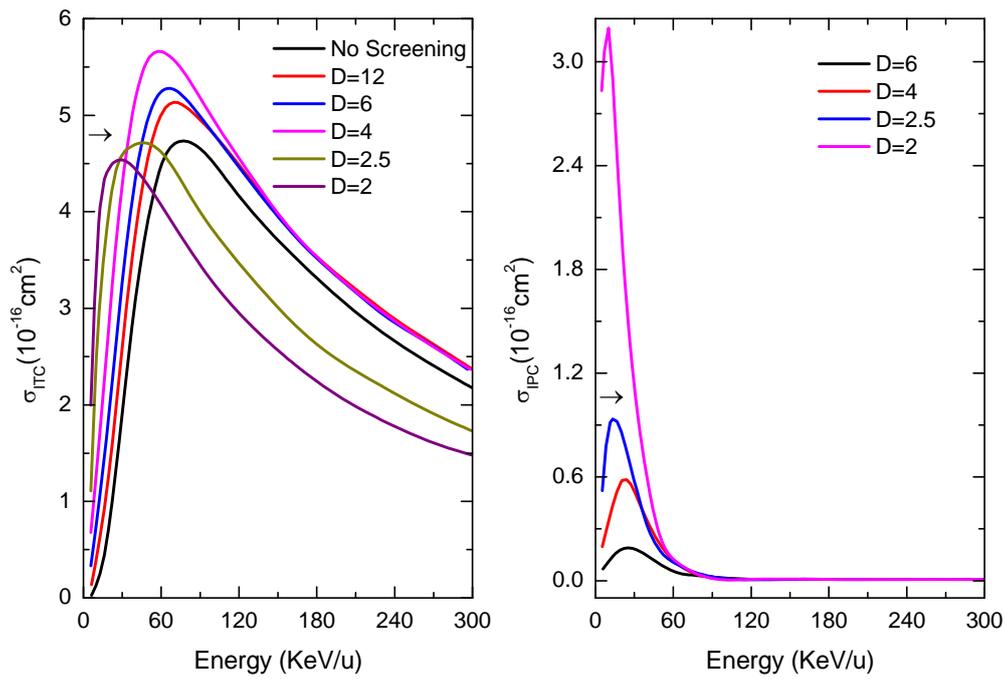

Fig. 8

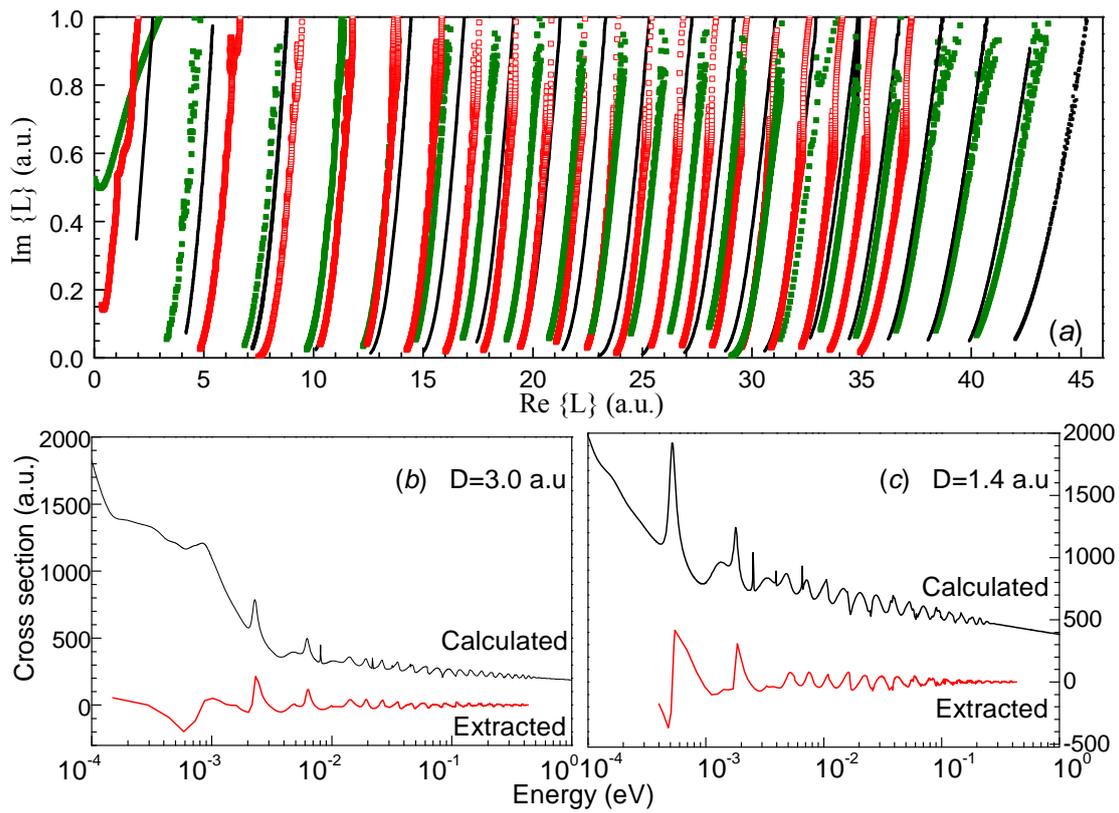

Fig. 9